\documentclass[aps,twocolumn,preprintnumbers,amsmath,amssymb,footinbib,prl]{revtex4}
\usepackage{graphicx}
\usepackage{bm}
\newcommand{\be}{\begin{equation}}
\newcommand{\ee}{\end{equation}}
\newcommand{\bea}{\begin{eqnarray}}
\newcommand{\eea}{\end{eqnarray}}

\newcommand{\lsim}{\mbox{\raisebox{-0.6ex}{$\stackrel{<}{\sim}$}}\:}

\begin{document}
\title{Inhomogeneous freeze-out in relativistic heavy-ion collisions}
\author{A.~Dumitru$^a$, L.~Portugal$^{a,b}$ and D.~Zschiesche$^a$}
\affiliation{$^a$Institut f\"ur Theoretische Physik, 
J.W.\ Goethe Universit\"at,\\
Max-von-Laue Str.\ 1, 60438 Frankfurt, Germany\\
$^b$Instituto de Fisica, Universidade Federal do Rio de Janeiro, Brazil}

\begin{abstract}
A QCD phase transition may reflect in a inhomogeneous decoupling
surface of hadrons produced in relativistic heavy-ion collisions.
We show that due to the non-linear dependence of the particle
densities on the temperature and baryon-chemical potential such
inhomogeneities should be visible even in the integrated,
inclusive abundances. We analyze experimental data from
Pb+Pb collisions at CERN-SPS and Au+Au collisions at BNL-RHIC to
determine the amplitude of inhomogeneities.
\end{abstract}

\maketitle

\noindent
\section{Introduction}
\label{intro}

Collisions of heavy nuclei at relativistic energies produce very hot and
baryon-dense QCD matter~\cite{HI}. The entropy per net baryon in the
central region increases monotonically with beam energy, i.e.\ the
maximum temperature increases, while the baryon-chemical potential
decreases. Hence, by varying the energy as well as the mass number of
the nuclei etc., one can explore various regimes of the phase diagram
of QCD. In particular, it is expected that at sufficiently high
energies a transient state of deconfined matter with broken $Z(3)$
center symmetry and/or with (approximately) restored chiral symmetry
is produced.

Lattice QCD simulations~\cite{Fodor:2001pe} indicate that a
second-order critical point exists, which was predicted by effective chiral
Lagrangians~\cite{SRS}; present estimates locate it at
$T\approx160$~MeV, $\mu_B\approx360$~MeV.  This point, where the
$\sigma$-field is massless, is commonly assumed to be the endpoint of
a line of first-order phase transitions in the $(\mu_B,T)$ plane.

There is an ongoing experimental effort to detect the line of
first-order phase transitions at high baryon density, and its possible
endpoint, in heavy-ion collisions. It is hoped that by varying the
beam energy, for example, one can ``switch'' between the regimes of
first-order phase transition and cross over, respectively.  If the
particles decouple shortly after the expansion trajectory crosses the
line of first order transitions one may expect a rather inhomogeneous
(energy-) density distribution on the freeze-out
surface~\cite{Bower:2001fq,Paech:2003fe} (similar, say, to the CMB
photon decoupling surface observed by WMAP~\cite{wmap}).  On the other
hand, if the low-temperature and high-temperature regimes are smoothly
connected, pressure gradients tend to wash out density
inhomogeneities. Similarly, in the absence of phase-transition induced
non-equilibrium effects, the predicted initial-state density
inhomogeneities~\cite{iniflucs,spherio} should be strongly damped.

Here, we investigate the properties of an inhomogeneous fireball at
(chemical) decoupling. Note that if the scale of these inhomogeneities
is much smaller than the decoupling volume then they can not be
resolved individually, nor will they give rise to large event-by-event
fluctuations. Because of the nonlinear dependence of the hadron
densities on $T$ and $\mu_B$, they should nevertheless reflect in the
{\em average} abundances. Our goal is to check whether the
experimental data shows any signs of inhomogeneities on the freeze-out
surface.

Our basic assumption is that as the fireball expands and cools, at
some stage the abundances of hadrons ``freeze'', keeping memory of the
last instant of chemical equilibrium. This stage is refered to as
chemical freeze-out. By definition, only processes that conserve
particle number for each species individually, or decays
of unstable particles may occur later on.

The simplest model is to treat the gas of hadrons within the grand
canonical ensemble, assuming a homogeneous decoupling volume.  The
abundances are then determined by two parameters, the temperature $T$
and the baryonic chemical potential $\mu_B$; the chemical potential
for strangeness is fixed by the condition for overall strangeness
neutrality, and other conserved charges shall be neglected. Fits of
hadronic ratios were performed extensively~\cite{Pbm05,thermo} 
within this
model, sometimes also including a strangeness ($\gamma_s$) 
or light quark ($\gamma_q$)  supression factor  
\cite{thermo_gammas,Becc05} or interactions with the chiral
condensate~\cite{thermo_int}.

The purpose of the present paper is to analyze the experimental data
on relative abundances of hadrons with respect to the presence of
inhomogeneities on the decoupling surface.  To that end, we propose a
very simple and rather schematic extension of the common grand
canonical freeze-out model, i.e.\ a {\em superposition} of such
ensembles with different temperatures and baryon-chemical
potentials. Each ensemble is supposed to describe the
local freeze-out on the scale of the correlation length $\sim 1/T\sim
1-2$~fm. Even if freeze-out occurs near the critical point, the
correlation length of the chiral condensate is bound from above by
finite size and finite time effects, effectively resulting in similar
numbers~\cite{corr_length}. On the other hand, for small chemical
potential, far from the region where the $\sigma$-field is critical,
the relevant scale might be set by the correlation length for Polyakov
loops, which is of comparable
magnitude~\cite{loop_corlength}. Classical nucleation theory for
strong first-order phase transitions predicts even larger
``bubbles''~\cite{nucleation} but is unlikely to apply to small,
rapidly expanding systems encountered in heavy-ion
collisions~\cite{Paech:2003fe,Scavenius:2000bb}. Another (classical)
model for the formation of small droplets in rapidly expanding QCD
matter has been introduced in~\cite{Mishustin:1998eq}.

The entire decoupling surface contains many such ``domains'', even if
a cut on mid-rapidity is performed. We therefore
expect that the distributions of temperature and chemical potential
are approximately Gaussian. Besides simplicity, another goal of the
present analysis is to avoid reference to a particular dynamical
model for the formation or for the distribution of density
perturbations. In fact, we presently aim merely at checking whether
any statistically significant signal for the presence of
inhomogeneities is found in the data. If so, more sophisticated
dynamical models could be employed in the future to understand the
evolution of inhomogeneities from their possible formation in a phase
transition until decoupling. 

Rate equations for nuclear fusion and dissociation processes (and
neutron diffusion) have been used for inhomogeneous big bang
nucleosynthesis in the early universe~\cite{iBBN}. Similarly, hadronic
cascade models could be used for heavy-ion
reactions~\cite{HydroCascade}. This would remove reference to the
grand-canonical ensemble and to a thin decoupling {\em surface} in
space-time. In fact, hadronic binary rescattering models do predict a
rather thick freeze-out layer~\cite{HydroCascade,sorge}, where matter
expands non-ideally. On the other hand, the steep drop of
multi-particle collision rates with temperature should narrow the
freeze-out again~\cite{BSW}. In either case, we do not expect a strong
energy dependence of the width of freeze-out (see also~\cite{ceres}).

At chemical freeze-out, matter is in a state of expansion. However,
such flow effects do not affect the relative abundances of the particles
(in full phase space) if their densities are homogeneous throughout
the decoupling volume. The total number of particles of species $i$,
integrated over a solid angle of $4\pi$, is given by an integral of
the current $N_i^\mu = \rho_i\,u^\mu$, with $u^\mu$ the four-velocity of the
expanding fluid, over a given freeze-out hypersurface $\sigma^\mu =
(t^{\rm fo},\vec{x}^{\rm \, fo})$: 
\be N_i = \int d\sigma_\mu N_i^\mu =
\rho_i(T^{\rm fo},\mu_B^{\rm fo}) \int  u^\mu d\sigma_\mu~.  
\ee 
The second factor on the r.h.s.\ is nothing but the three-volume $V_3$
of the decoupling hypersurface as seen by the observer. This volume is
common to all species and drops out of multiplicity ratios: $N_i/N_j =
\rho_i^{\rm fo}/\rho_j^{\rm fo}$. It is clear that the argument holds
even when cuts in momentum space are performed, provided that the
differential distributions of all particles do not depend on that
particular momentum-space variable (for example, rapidity cuts for
boostinvariant expansion~\cite{CleyRed99}).

When the intensive variables $T$ and $\mu_B$ vary,
then the integration measure $(\int u\cdot d\sigma)/V_3$
will, in general, depend on the assumed distribution and amplitude of
inhomogeneities, as well as on the hydrodynamic flow profile etc.\
Nevertheless, it is still the same for all particle species and so can
be written in the form
\be
{1\over V_3}\int u\cdot d\sigma \longrightarrow \int dT d\mu_B \, P(T,\mu_B)~,
\ee
with $P(T,\mu_B)$ some distribution for $T$ and $\mu_B$. For
simplicity, and for lack of an obvious motivation for assuming
otherwise, we shall take $P(T,\mu_B)$ to factorize into a distribution
for $T$, times one for $\mu_B$. These distributions could, in
principle, be obtained from the real-time evolution of the phase
transition~\cite{Bower:2001fq,Paech:2003fe}.

\section{The model}  \label{model}
With these qualifications aside, we now proceed to introduce our model and
to analyze the available data from heavy-ion collisions at high energies.
The hadron abundances are determined by four parameters: the
arithmetic means of the temperatures and chemical potentials of all domains,
$\overline{T}$ and $\overline{\mu}_B$, and the widths of their
Gaussian distributions, $\delta T$ and $\delta \mu_B$.
Then, the average density of species $i$ is computed as 
\bea \label{ave_dens}
& &\overline{\rho}_i\; (\overline{T},\overline{\mu}_B, \delta
T,\delta\mu_B) = \\
& &\int\limits_0^\infty dT \;
P(T;\overline{T},\delta T) \int\limits_{-\infty}^\infty d\mu_B \;
P(\mu_B; \overline{\mu}_B,\delta\mu_B)~\rho_i (T,\mu_B)~, \nonumber
\eea
with $\rho_i(T,\mu_B)$ the actual ``local'' density of species
$i$, and with $P(x; \overline{x},\delta x) \sim \exp
[-{\left(x-\overline{x}\right)^2}/{2\, \delta x^2} ]$
the distribution of temperatures and chemical potentials within the
decoupling three-volume (the proportionality constants
normalize the distributions over the intervals where they are
defined). In the limit $\delta T$, $\delta\mu_B\to0$
the Gaussian distributions are replaced by $\delta$-functions and the
conventional homogeneous freeze-out scenario is recovered:
$\overline{\rho}_i\; (\overline{T},\overline{\mu}_B, 0,0) =
\rho_i(\overline{T},\overline{\mu}_B)$. In other words, in that limit
the average densities are uniquely determined by the first moments of
$T$ and $\mu_B$.

For the present analysis we compute the densities $\rho_i (T,\mu_B)$
in the ideal gas approximation, supplemented by an ``excluded volume''
correction:
\be \label{DensityExclVol}
\rho_i (T,\mu_B) = \frac{\rho_i^{\rm id-gas} (T,\mu_B)}{1+ 
v_i \sum_j \rho_j^{\rm id-gas} }~.
\ee
This schematic correction models repulsive interactions among the
hadrons at high densities. $v_i$ denotes the volume occupied by a
hadron of species $i$; we employ $v=\frac {4}{3} \pi {R_0}^3$ with
$R_0= 0.3$~fm for all species~\cite{ExclVol}. Therefore, for the homogeneous
model the denominator in~(\ref{DensityExclVol}) drops out of
multiplicity ratios. This is not the case for an inhomogeneous
decoupling surface, where the distributions of various species differ.

The densities of strange particles depend also on the
strangeness-chemical potential $\mu_S$, which we determine by imposing
local strangeness neutrality. Strictly speaking, this condition is not
mandatory, of course. The effect of independent fluctuations of
$\mu_S$ should be looked at in the future, in particular for
collisions at low and intermediate energies ($\surd s_{NN} \lsim
15$~GeV). This may help to reproduce the $\overline\Lambda$ to
$\overline{p}$ ratio, which was found to be larger
than one~\cite{inh_mus} and the $K^+/\pi^+$ enhancement around
$E_{\rm{Lab}}/A=30$ GeV~\cite{Na49_data}. 
For all fits over the full solid angle, we
fixed the isospin chemical potential by equating the total charge in
the initial and final states; for the mid-rapidity fits at high
energies, we fixed $\mu_I=0$.

To illustrate the effect of inhomogeneities on the distributions of
various hadrons within the decoupling volume we introduce
\bea
& & D_i(T;\overline{T},\overline{\mu}_B, \delta T,\delta\mu_B)
      = P(T;\overline{T},\delta T)\nonumber \\ & &\hspace*{1cm}\times
\frac{\int\limits_{-\infty}^\infty d\mu_B \; P(\mu_B;
\overline{\mu}_B,\delta\mu_B)~\rho_i (T,\mu_B)~}
{\overline{\rho}_i\; (\overline{T},\overline{\mu}_B, \delta
  T,\delta\mu_B)}~, \label{D_T}  \\
& &\nonumber \\
& & D_i(\mu_B;\overline{T},\overline{\mu}_B, \delta T,\delta\mu_B)
      =P(\mu_B;\overline{\mu}_B,\delta\mu_B)\nonumber \\ & & \hspace*{1cm}\times
\frac{\int\limits_0^\infty 
dT \;P(T;\overline{T},\delta T)~\rho_i (T,\mu_B)~}
{\overline{\rho}_i\; (\overline{T},\overline{\mu}_B, \delta
  T,\delta\mu_B)}~.\label{D_mu}
\eea
$D_i(T)$, for example, is the probability that a particle of type $i$ 
was emitted from a domain of temperature $T$. 
The main contribution to the integrals in~(\ref{ave_dens}) is {\em
  not} from $\overline{T}$ and $\overline{\mu}_B$ since hot spots
shine brighter than ``voids''. Rather, they are
dominated by the stationary points of the distributions
defined in eqs.~(\ref{D_T},\ref{D_mu}) above. Hence, the average
emission temperature $\langle
T\rangle_i$ and baryon-chemical potential $\langle\mu_B\rangle_i$
in general depend on the particle species $i$, unless $\delta
T=\delta\mu_B=0$. They can be evaluated as
\bea
\langle T\rangle_i &=& \int\limits_0^\infty 
    dT \;T\;  D_i(T;\overline{T},\overline{\mu}_B,
    \delta T,\delta\mu_B)~, \nonumber\\
\langle\mu_B\rangle_i &=& \int\limits_{-\infty}^\infty d\mu_B \;\mu_B\;
D_{i}(\mu_B;\overline{T},\overline{\mu}_B, \delta T,\delta\mu_B)~.
\label{averages}
\eea
Physically, this means that for non-zero widths of the temperature and
chemical potential distributions the freeze-out volume is not
perfectly ``stirred'', in that the relative concentrations of the
particles vary.

For some limiting cases one can estimate the effect analytically. Consider
first massless particles without chemical potential ($\approx$ direct
pions):
\be
 D(T;\overline{T}, \delta T) \sim \exp \left(3\log
 \frac{T}{\overline T}
 -\frac{\left(T-\overline{T}\right)^2}{2\; \delta T^2} \right)~.
\ee
To leading order in $\delta T/\overline{T}$ the stationary point of
the exponential is
\be \label{avTlight}
\frac{T^*}{\overline{T}} = 1 + 3 \frac{\delta
  T^2}{\overline{T}^2} + \cdots
\ee
Hence, massless particles are not very sensitive to small fluctuations in
temperature. They are typically emitted from regions with temperature
$\approx \overline{T}$, up to quadratic corrections in
${\delta T}/{\overline{T}}$. In other words,
the ``particle emission distribution" $D(T)$ is shifted (and skewed)
only slightly from the ``temperature distribution" $P(T)$.
This, of course, is simply due to the $\rho \sim T^3$ power-law form of the 
local density of massless particles. The convolution of a (narrow) Gaussian 
with a power-law does not lead to a large shift of the peak.

Next, we consider the more interesting case of massive particles with
 chemical potential equal to $\mu_B$ (for anti-baryons replace
 $\overline{\mu}_B\to -\overline{\mu}_B$ etc.). We assume that
 $\exp[ (\mu_B-m)/T] \ll 1$ and $m/ T\gg1$ such that
quantum statistical and relativistic corrections can be neglected, but
 allow that
$x\equiv(m/\overline{T})(\delta T^2/\overline{T}^2) = {\cal O}(1)$.
The integral over $\mu_B$ is then straightforward, leading to the 
particle distribution
\bea
& & D(T)\;\sim\; \nonumber\\
& & \exp \left ( \frac{3}{2} \log (T/\overline{T}) + \frac{\mu_B - m}{T} + 
\frac{{\delta\mu_B}^2}{2{\overline{T}}^2} -
 \frac{\left(T-\overline{T}\right)^2}{2\; \delta T^2} \right )~. 
 \label{DT_heavy}
\eea
Again, we look for the stationary point of the 
exponential. To leading order in the Gaussian width,
\bea
\frac{T^*}{\overline{T}} &=& 
           1 + \frac{{\delta T}^2}{{\overline{T}}^2}
\left ( \frac{m}{ \overline{T}} 
 - \frac{\mu_B}{ \overline{T}} + \frac{3}{2} \right) \nonumber\\
&=& 1 + \frac{{\delta T}^2}{{\overline{T}}^2} \left ( \frac{m}{
   \overline{T}} + O(1) \right)  \simeq 1+x~. \label{avTheavy}
\eea
In the second step, we have used the fact that in the Boltzmann limit
${\mu_B}/{ \overline{T}}$ is of order 1 (i.e., not
parametrically large). Hence, for massive particles the distribution
is shifted by a large amount $x={\cal O}(1)$. Their emission is
dominated by the tails of the Gaussian distribution or, in physical
terms, by rare ``hot spots''. 

To estimate the increase of the average density relative to the
homogeneous case we plug the expression for $T^*$ into
$D(T)$ from eq.~(\ref{DT_heavy}), which gives 
\bea
& &\frac{\overline{\rho}(\overline{T},\delta T, \overline{\mu}_B, \delta
  \mu_B)}{\rho(\overline{T},\overline{\mu}_B)}= \nonumber \\
& &\hspace*{1cm}
  f\, \exp \left ( \frac{1}{(1+x)^2} \,\frac {{\delta\mu_B}^2}{2\,
  {\overline{T}}^2} +
\frac{x}{1+x} \,\frac {  m-\overline{\mu}_B}{\overline{T}} \right)~.
\label{densinc}
\eea
The non-exponential prefactor $f$ arises due to the changing width of
the integration measure when the temperature is inhomogeneous
and can not be estimated by
a saddle-point integration ($f=1$ for $\delta T=0$). Regardless, the
main issue here is that this ratio increases exponentially
with $(\delta\mu_B/\overline{T})^2$.
Hence, we conclude that small fluctuations in the chemical potential are 
sufficient to raise the number of heavy particles significantly. Note
that what matters is the magnitude of $\delta\mu_B$ relative to
$\overline{T}$, not $\overline{\mu}_B$.

Similarly, temperature inhomogeneities also increase the
density exponentially.  That is, for $\delta T \rightarrow 0$,
$x/(1+x)\approx x$. However, the growth saturates when $x\sim 1$. 
The enhancement factor from $x=0$ to $x\sim 1$ is very large, $\sim
\surd\exp \left( {(m-\overline{\mu}_B)}/{\overline{T}} \right) $.

\section{Data analysis}

To determine the four parameters of the model we minimize
\begin{equation} \label{chi2}
\chi^2 = \sum_i {\left(r_i^{exp} - r_i^{model}\right)^2}/{\sigma_i^2}
\end{equation}
in the space of $\overline{T}$, $\overline{\mu}_B$, $\delta T$, and
$\delta\mu_B$. That is, we obtain least-square estimates for the
parameters, assuming that they are independent.
In~(\ref{chi2}), $r_i^{exp}$ and
$r_i^{model}$ denote the experimentally measured and the calculated
particle ratios, respectively, and $\sigma_i^2$ is set by the
uncertainty of the measurement.  Wherever available, we sum systematic
and statistical errors in quadrature.

The data used in our analysis are the particle multiplicities measured
by the NA49 collaboration for central Pb+Pb collisions at beam energy
$E_{\rm Lab}/A=20$, 30, 40, 80 and 158~GeV~\cite{Na49_data}, and those
measured by STAR for central Au+Au collisions at BNL-RHIC,
ref.~\cite{RHIC130_data} 
($\surd{s_{NN}}= 130$~GeV, compiled in \cite{RHIC_130_comp})
and ref.~\cite{RHIC_data} (200~GeV).
At RHIC energies, we analyze the midrapidity data; at top SPS energy,
both, midrapidity and $4\pi$ data.  At all other energies, we restrict
ourselves to the $4\pi$ solid angle data by NA49 in order to avoid
biases arising from differing acceptance windows of various
experiments. Furthermore, our checks showed that the fit results can
depend somewhat on the actual selection of experimental ratios. Hence,
where possible, we have opted for the least bias by choosing $r_i^{exp}
= N^{exp}_i / N^{exp}_\pi$, i.e.\ the multiplicity of species $i$
relative to that of pions. This represents the maximal set of
independent data points, as it is equivalent to fitting {\em absolute
multiplicities} with an additional overall three-volume parameter,
$N_i = V_3 \rho_i$.

Specifically, at $E_{\rm Lab}/A=20$, 30, and 80~GeV the multiplicities
of $\pi^+$, $\pi^-$, $K^+$, $K^-$, $B-\overline{B}$, $\Lambda$,
$\overline\Lambda$, and $\phi$ are available. For the (in-)homogeneous
model, this leaves five (three) degrees of freedom. At 40~GeV, we can add
the $\Xi^-$ and $\Omega+\overline{\Omega}$.  The data sets for
top SPS energies include yet a few more species: $p$, $\overline{p}$ (only
midrapidity), $K^0_S$ (only $4\pi$), $\overline\Xi^+$ and $\Omega$,
$\overline\Omega$ seperately. For RHIC-130, we fitted to the
$K^+/K^-$, $\overline{p}/p$, $\overline{\Lambda}/\Lambda$,
$\Xi^+/\Xi^-$, $\overline{\Omega}/\Omega$, $K^-/\pi^-$, $K^0_S/\pi^-$,
$\overline{p}/\pi^-$, $\Lambda/\pi^-$, $K_0^{\ast}/\pi^-$,
$\phi/\pi^-$, $\Xi^-/\pi^-$ and $\Omega/\pi^-$ ratios.  Finally, at RHIC-200
the $K^+/K^-$, $\overline{p}/p$,
$\overline{\Omega}/\Omega$, $K^-/\pi^-$, $\overline{p}/\pi^-$,
$\Lambda/\pi^-$, $\overline{\Lambda}/\pi^-$, $\Xi^-/\pi^-$,
$\Xi^+/\pi^-$, $\Omega/\pi^-$, $\phi/K^-$ and $K_0^{\ast}/K^-$ ratios
were used. The first three ratios are close to unity and essentially
just set the chemical potentials to zero; they do not help to fix
$\overline{T}$, $\delta T$ and $\delta\mu_B$.

Where appropriate, feeding from strong and electromagnetic decays has
been included in $r^{model}_i$ by replacing ${\rho}_i
\rightarrow {\rho}_i + B_{ij}\; {\rho}_j.$ The
implicit sum over $j\neq i$ runs over all unstable hadron species,
with $B_{ij}$ the branching ratio for the decay $j\to i$, which were
taken from \cite{PDG}. From all the resonances listed by the Particle
Data Group \cite{PDG}, mesons up to a mass of 1.5~GeV and baryons up
to a mass of 2~GeV were included, respectively.  The finite widths of
the resonances were not taken into account, and unknown branching
ratios were excluded from the feeding.  These details are irrelevant
for the qualitative behavior of $\delta T$ and $\delta \mu_B$ but do,
of course, matter for quantitative results.  For SPS energies,
feeding from weak decays was included only for the ${\Xi} \to\Lambda$
and $\Omega \to\Lambda$ channels (and the corresponding decays of the
anti-baryons) at $E_{\rm Lab}/A=20$ and 30~GeV. All other
experimental yields were already corrected by NA49 for feed-down from
weak decays~\cite{Na49_data}.  At RHIC energies, feeding from weak
decays has to be taken into account for a variety of particle species,
c.f.~\cite{RHIC130_data,RHIC_data}.

Technically, our fits use a four-dimensional lookup table for
$\overline{\rho}_i$ with 1~MeV steps in $T$ and $\delta T$ and $10$
MeV steps in $\mu$ and $\delta \mu$ for the inhomogeneous fits, and 1
MeV steps for both $T$ and $\mu$ for the homogeneous fits.  The finite
grid of course limits our ability to determe the best fit. However,
the effect was checked to be small, so that the accuracy should be
sufficient to investigate the qualitative behavior of inhomogeneities,
and whether they can improve the agreement with the experimental data
significantly.

\section{Results}

\begin{figure}[h]
\vspace{-1.0cm}
\includegraphics[width=9cm]{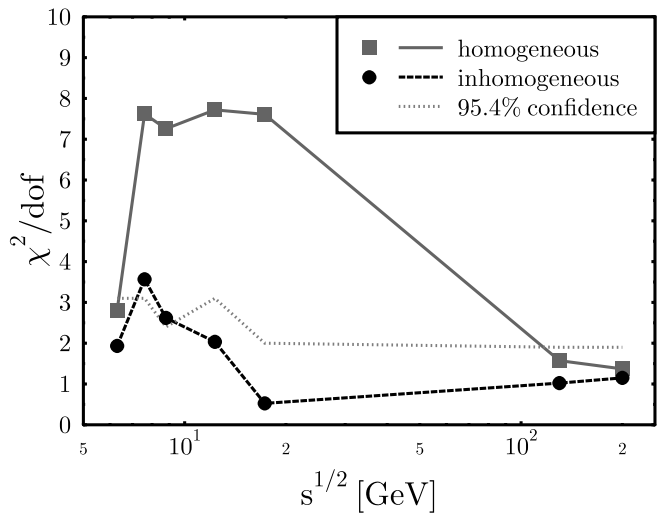}
\vspace{-0.6cm}
\caption{\label{chi2vssqrts}
$\chi^2/dof$ versus $\surd{s_{NN}}$ for the homogeneous 
($\delta T = \delta \mu = 0$, squares) and the inhomogeneous fit 
($\delta T$ and $\delta \mu$ free parameters, circles). 
The full and dashed lines are meant to guide the eye. Furthermore,
the $\chi^2/dof$ corresponding to the $95.4 \%$ confidence interval
is shown by the dotted line.}
\end{figure}
Fig.~\ref{chi2vssqrts} shows the minimal $\chi^2$ per degree
of freedom (taken as the number of data points minus the number of
parameters) for the homogeneous and the inhomogeneous approach,
respectively. 
At $E_{\rm Lab}/A=20$~GeV and at RHIC energies,
$\chi^2/dof$ is similar for both models. Thus, the inhomogeneous
model does not provide a statistically significant improvement of the
description of the measured particle ratios.
Hence, the assumption of a
nearly homogeneous decoupling surface can not be rejected for
low SPS and RHIC energies.

On the other hand, for intermediate SPS energies, $E_{\rm Lab}/A
\simeq 30-160$~GeV, $\chi^2/dof$ is considerably smaller for the
inhomogeneous freeze-out surface than for the homogeneous case, which
is far outside the 95.4\% confidence interval \cite{spsfits}.
This indicates that the
parameters of the inhomogeneous model are well determined and have
reasonably small error bars. It is worth noting that, in general, the
improvement is not driven by one single species; rather, the
inhomogeneous model describes nearly all multiplicities better than a
homogeneous decoupling surface~\cite{Bormio}. We also stress that
our fits with $\delta T=\delta\mu_B=0$ reproduce results from the
literature~\cite{thermo,thermo_gammas,RHIC_data} if the same input
ratios are selected.

\begin{figure}
\vspace{-1.0cm}
\includegraphics[width=9cm]{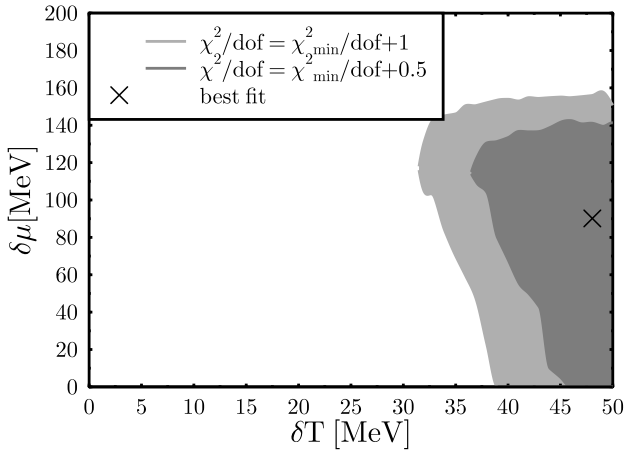}
\vspace{-0.6cm}
\caption{\label{dtdmu_cor_sps158}
$\chi^2/dof$ contours in the $\delta T$, $\delta\mu_B$ plane for top
  SPS energy, $E_{\rm Lab}=158$~GeV. The other two parameters
  ($\bar{T}$, $\bar{\mu}_B$) are allowed to vary freely. The
  $\chi^2/dof$ minimum is indicated by the cross.}
\end{figure}
\begin{figure}
\vspace{-1.0cm}
\includegraphics[width=9cm]{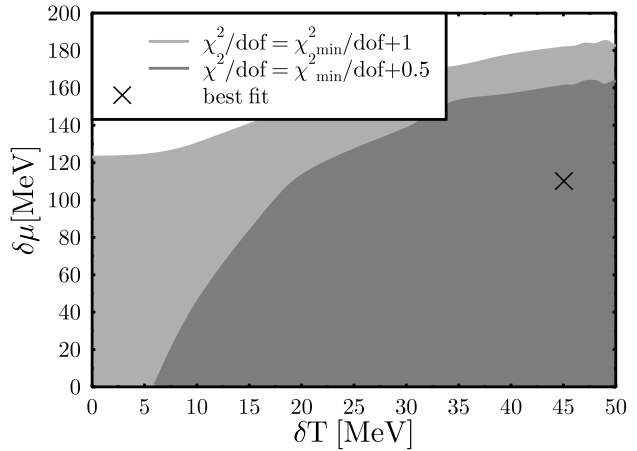}
\vspace{-0.6cm}
\caption{\label{dtdmu_cor_rhic200}
Same as fig.~\ref{dtdmu_cor_sps158} for RHIC energy
($\surd{s_{NN}}=200$~GeV).}
\end{figure}

To illustrate the significance of inhomogeneities differently, we show
contours of $\chi^2/dof$ in the plane of $\delta T$, $\delta\mu_B$ in
figs.~\ref{dtdmu_cor_sps158} and~\ref{dtdmu_cor_rhic200}. Here,
$\overline{T}$ and $\bar{\mu}_B$ were allowed to vary freely such as
to minimize $\chi^2$ at each point. Fig.~\ref{dtdmu_cor_sps158} shows
that $\chi^2$ is relatively flat along the
$\delta\mu_B$ direction, while $\delta T$ is determined more
accurately and is clearly non-zero. In general we find that there is 
little correlation between $\delta T$ and $\delta\mu_B$ and that 
about the minimum, $\chi^2$ is rather flat in $\delta\mu_B$ direction.

On the other hand, Fig.~\ref{dtdmu_cor_rhic200} shows that at RHIC
energy, $\chi^2$ is very flat in both directions. With the present
data points, a homogeneous freeze-out model appears to be a reasonable
approximation at high energies.  We have traced the origin of the very
flat $\chi^2(\delta T)$ at RHIC energy to a strong anti-correlation between
$\delta T$ and $\overline{T}$. This degeneracy appears, in the regime
of small chemical potentials, when the data is dominated by particles
of similar mass (including contributions from resonance
decays). There is then no distinguishing feature which would fix both
$\delta T$ and $\overline{T}$ independently, i.e.\ larger $\delta T$
can be traded for smaller $\overline{T}$ and vice versa.
Eq.~(\ref{densinc}) indicates that the
model requires several species whose densities are dominated
by significantly different mass scales.

We now proceed to analyze the energy dependence of the freeze-out
conditions. As already mentioned above, the fit parameters
$\overline{T}$, $\overline{\mu}_B$, $\delta T$ and $\delta\mu_B$ do
not acquire a direct physical meaning; for example, $\overline{T}$ is
simply the arithmetic mean of the temperature within the entire volume,
but particle production is dominated by ``hot spots''. Similarly,
the RMS variation of the temperature within the decoupling volume,
$\delta T$, is not a direct measure for the different emission
temperatures of various particles, since the latter are given by a
{\em convolution} of the Gaussian temperature distribution with the
respective particle densities $\rho_i$, and so depend also on the
particle masses, the chemical potentials etc.

Hence, rather than discussing the energy dependence of the above
technical parameters, we instead focus on the particle emission
temperatures $\langle T\rangle_i$ and baryon-chemical potentials
$\langle \mu_B\rangle_i$ introduced in~(\ref{averages}). For each
energy, we define an average particle emission temperature by
summing over all species $i$ ({\em after} resonance decays):
\be \label{aveTparticles}
\overline{\langle T\rangle} = \frac{\sum_i \langle T\rangle_i
\, \overline{\rho}_i}{\sum_i \overline{\rho}_i}~, 
\ee
and similarly for $\overline{\langle \mu_B\rangle}$.

Along the same lines, we can determine the variance of the emission
temperatures (and chemical potentials) of various particle species as
\be \label{delTparticles}
\delta \langle T\rangle = \sqrt{\frac{\sum_i \left(\langle T\rangle_i -
  \overline{\langle T\rangle} \right)^2 \, \overline{\rho}_i}
{\sum_i \overline{\rho}_i}}~.
\ee
We repeat that all of these quantities are entirely determined by the
parameters $\overline{T}$, $\overline{\mu}_B$, $\delta T$ and
$\delta\mu_B$ characterizing the Gaussian distributions of temperature
and baryon-chemical potential; they do not represent additional fit
parameters. In particular, when $\delta
T = \delta\mu_B=0$, the homogeneous model is recovered:
$\overline{\langle T\rangle} = T^{fo}$, $\delta \langle T\rangle = 0$ etc.

\begin{figure}
\includegraphics[width=8cm]{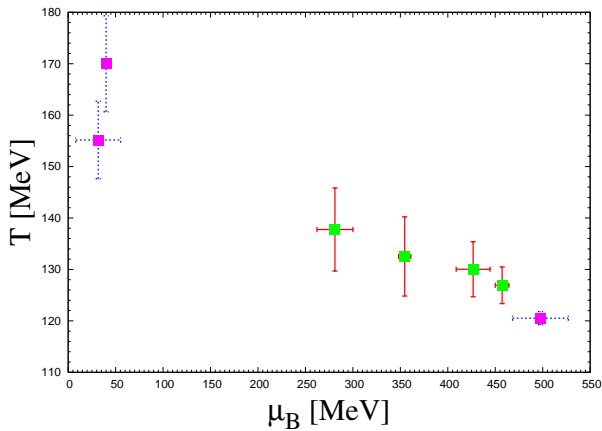}
\caption{\label{focurve}
Average particle emission temperature $\overline{\langle T\rangle}$
and chemical potential $\overline{\langle \mu_B\rangle}$ at
$E_{\rm Lab}/A=20$, 30, 40, 80, 158~GeV (CERN-SPS) and $\surd
s_{NN}=130$ and 200~GeV (BNL-RHIC), from right to left (symbols). The ``error
bars'' indicate the RMS deviations over particle species (i.e.\
$\delta \langle T\rangle$ and $\delta \langle \mu_B\rangle$).
For SPS-20 and RHIC, these are depicted by dotted 
lines since the inhomogeneities are not statistically significant.
}
\end{figure}
Fig.~\ref{focurve} depicts the energy dependence of the average
freeze-out temperature and chemical potential, as defined in
eq.~(\ref{aveTparticles}). The general trend is the same as in
the homogeneous model: $\overline{\langle
  T\rangle}$ increases with energy, while $\overline{\langle
  \mu_B\rangle}$ decreases. Their values agree to within $\sim15\%$
with those for a homogeneous system obtained previously by others
(cf.\ for example fig.~3 in Redlich {\it et al.} or
fig.~27 in Braun-Munzinger {\it et al.}~\cite{thermo}).
The inhomogeneous model, however, predicts sizeable variations of the
emission temperatures of different particle species: this is indicated
in fig.~\ref{focurve} by the ``error bars'', which correspond to
$\delta \langle T\rangle$ and $\delta \langle \mu_B\rangle$ as defined
in eq.~(\ref{delTparticles}). They were obtained from the best fit at
each energy, i.e.\ using the parameters corresponding to the lowest
$\chi^2/dof$ from the inhomogeneous model. We repeat, however, that at
the lowest and highest energies
the inhomogeneous fit has no statistical significance, in
that $\chi^2/dof$ is only marginally better than for the homogeneous
model. Hence, at these energies, the resulting $\delta \langle
T\rangle$ and $\delta \langle \mu_B\rangle$ 
(dotted lines) should be taken with $\sim100\%$ error.

At intermediate and high SPS energies, however, the variance of the
temperature can be determined reasonably well and is significantly
larger than zero.  $\delta\langle T\rangle$ appears to be rather large
already at $E_{\rm Lab}/A=30$~GeV, and increases further towards
higher SPS energies. As already stated, $\chi^2(\delta \mu_B)$ is
rather flat for all energies, so that $\delta \langle \mu_B\rangle$
should also be taken with some uncertainty.

\begin{figure}
\includegraphics[width=9cm]{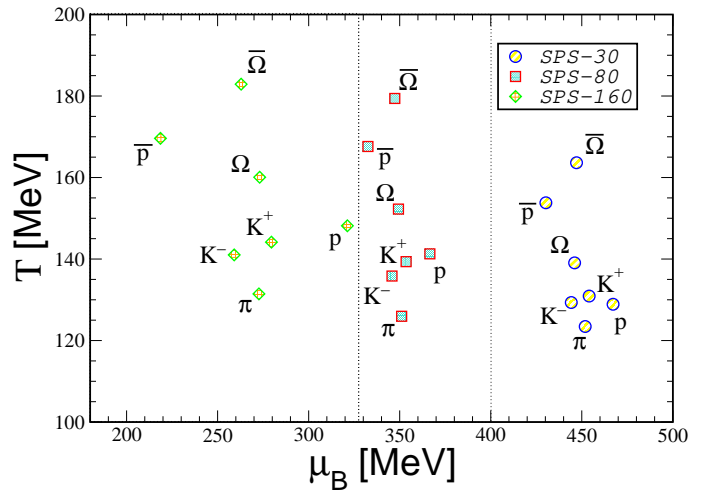}
\vspace{-0.6cm}
\caption{\label{foparticles}
Freeze-out temperatures $\langle T\rangle_i$
and chemical potentials $\langle \mu_B\rangle_i$ of various particle
species at $E_{\rm Lab}/A=30$, 80, 158~GeV.}
\end{figure}
Fig.~\ref{foparticles} shows the freeze-out temperatures and chemical
potentials for individual particle species at selected energies in the
CERN-SPS range. The effect of the inhomogeneities is evident. For
example, anti-protons are typically emitted from regions with
lower baryon-chemical potential than protons; also, heavy particles
are concentrated in ``hot spots'' while light pions are
distributed more evenly throughout the decoupling volume etc.
The temperatures of rare particle species can deviate much more from
$\overline{\langle T\rangle}$ than $\delta\langle T\rangle$
from fig.~\ref{focurve} might suggest. This is, of course, due to their
small weight in eq.~(\ref{delTparticles}).
Finally, we note that the rather high temperatures of the hot spots
from which the heavy particles emerge might indicate the need for a better
treatment of interactions~\cite{thermo_int} than the simple
excluded-volume model employed here.

\section{Summary and Outlook}
We have shown that inhomogeneities on the freeze-out hypersurface do
not average out but reflect in the {\em 4$\pi$ (or midrapidity),
  single-inclusive} abundances of
various particle species. This is due to the non-linear dependence of
the hadron densities $\rho_i(T,\mu_B)$ on the local temperature and
baryon-chemical potential. Consequently, even the average
$\overline{\rho}_i$ probe higher moments of the $T$ and $\mu_B$
distributions. Searching for such inhomogeneities could represent a
promising observable for a first-order phase transition.  

We introduced a simple model for the statistical description of
chemical freeze-out, extending the commonly used grand-canonical
ensemble to the case of inhomogeneous decoupling surfaces.  If
freeze-out occurs shortly after a first-order phase transition,
hydrodynamic flow and diffusion~\cite{Bower:2001fq,diff} may not
completely wash out inhomogeneities produced in the course of the
phase transition. Given that the number of such ``domains'' is
expected to be large, the inclusive distribution of temperature and
baryon-chemical potential on the decoupling surface should be
approximately Gaussian.

The model improves the fits of the measured particle ratios in the
CERN-SPS energy regime $E_{\rm Lab}/A\simeq30-160$~GeV
significantly. Homogeneous freeze-out is well outside the 95.4\%
confidence interval. This suggests that at intermediate energies the
freeze-out surface is not well ``stirred''.  Future studies could
perhaps shed more light on whether these inhomogeneities can indeed be
interpreted as fingerprints of a first-order phase
transition. Eventually, one would want to establish more quantitative
relations between the amplitudes of the $T$, $\mu_B$ inhomogeneities
and the properties of the phase transition, e.g.\ its latent heat and
interface tension.  Furthermore, the role of inhomogeneities in the
net-strangeness distribution should be studied.  On the other hand,
within the present model, no statistically significant improvement
over homogeneous freeze-out was observed at lower (SPS-20) and higher
(RHIC) energies. 

Inhomogeneities could also affect the coalescence probabilities of
(anti-) nucleons to light (anti-) nuclei, which are also sensitive to
density perturbations~\cite{ioffe}. Other signals, such as two-particle
correlations~\cite{spherio,bubble}, could also be analyzed in this regard.

To improve the quality of the statistical fits, more data on hadron
multiplicities would be helpful, in particular at the lower end of the
CERN-SPS energy spectrum and at RHIC. This includes estimates of
multiplicities of unstable resonances ($\rho$, $K^*$, $\omega$,
$\Delta$ ...) at chemical freeze-out~\cite{reson}.  Data from GSI-FAIR
and CERN-LHC will provide additional constraints for the evolution of
chemical freeze-out with energy.

\vspace*{.5cm}
{\bf Note added:} While we finished this manuscript, the
superstatistics approach~\cite{SuperStat} was
brought to our attention. It considers non-equilibrium systems in
stationary states with fluctuating intensive quantities, which are
described by a superposition of Boltzmann ensembles. Although our
approach emerged from a different physical picture, it can
nevertheless be viewed as an application of ``superstatistics'' to
particle freeze-out in heavy-ion collisions.

\vspace*{1cm} 
{\centerline{\em{\bf Acknowledgements}}} 
We thank C.~Greiner for fruitful remarks concerning the model,
C.~Blume and M.~Gazdzicki for helpful discussions about the NA49 data and
A.~Grunfeld for helping with the construction of the resonance table.
L.P.~thanks CAPES for supporting a
one-year stay at ITP, Goethe University.  A.D.\ and D.Z.\ acknowledge the
hospitality of the Nuclear Theory group during a stay at UFRJ
sponsored by DAAD (PROBRAL program), where this work was completed.
This work used computational resources provided by the Center for Scientific 
Computing (CSC) of Goethe University, Frankfurt.


\begin{thebibliography}{99}
\bibitem{inh_mus} C.~H\"ohne (for the NA49 collaboration), nucl-ex/0510049.
Our present best fit (of the $4\pi$ ratios) for the inhomogeneous
model without independent fluctuations of $\mu_S$ yields
$\overline\Lambda/\overline{p}=0.93$ at $E_{\rm Lab}/A=40$~GeV, for
example, versus 0.76 for $\delta T=\delta\mu_B=0$ (without
contributions from weak decays).
\bibitem{spsfits}
For SPS-158 only the
$4\pi$ fit is shown; restricting the homogeneous fit to the
mid-rapidity data gives smaller $\chi^2/dof$, but still significantly
higher than in the inhomogeneous approach. 
$\chi^2$ is smaller if other particle ratios are considered, as 
for example $\Xi/\Lambda, \Omega/\Xi$ instead of $\Xi/\pi, \Omega/\pi$
\cite{Pbm05} or if finite widths of resonances are taken into 
account \cite{Becc05}. However, the increase of $\chi^2$ 
at SPS energies is generic if Na49 $4\pi$-data are fitted.
\end{thebibliography}
\end{document}